\documentclass[12pt]{article}
\usepackage{amssymb}
\usepackage{amsmath}
\usepackage{hyperref}
\usepackage[dvips]{graphicx}
\usepackage{bm}
\usepackage{latexsym}
\begin{document}
%	\linenumbers  
\title{\bf   Nonlinear Yang-Mills AdS black brane and DC conductivity }
\author{Mehdi Sadeghi\thanks{Corresponding author: Email: mehdi.sadeghi@abru.ac.ir} \hspace{2mm}\\\\
	{\small {\em  Department of Physics, Faculty of Basic Sciences,}}\\
{\small {\em  Ayatollah Boroujerdi University, Boroujerd, Iran}}
}
\date{\today}
\maketitle

\abstract{ In this paper, we examine Einstein-Hilbert gravity featuring a cosmological constant and a non-abelian nonlinear electromagnetic field that is minimally coupled to gravity. We first present the black brane solution for this model and subsequently calculate the color non-abelian DC conductivity for this solution using AdS/CFT duality. Our results retrieve the Yang-Mills model in the limit as $q_1$ approaches zero.}\\

%\pacs{11.10.Jj, 11.10.Wx, 11.15.Pg, 11.25.Tq}


%--------------------------------------------------------------------------
\section{Introduction} \label{intro}

Maxwell’s electromagnetic field theory, being inherently linear, experiences several significant challenges\cite{Delphenich:2003yw,Delphenich:2006ec,EslamPanah:2021xaf}, including singularities at the particle's origin, infinite self-energy issues, and complications arising from the self-interaction of virtual electron-positron pairs\cite{W. Heisenberg_1936}. Additionally, it cannot adequately describe radiation propagation in specific materials\cite{DeLorenci:2001gf}. Heisenberg and Euler\cite{W. Heisenberg_1936} demonstrated that quantum electrodynamics (QED), through loop corrections, can lead to a framework of nonlinear electrodynamics (NED). The combination of NED with general relativity (GR) offers potential explanations for cosmic phenomena such as cosmic inflation and the current accelerated expansion of the Universe\cite{Camara:2004ap}-\cite{Garcia-Salcedo:2013cxa}.\\
Various theories of nonlinear electrodynamics have emerged, including logarithmic gauge theory\cite{Gaete:2013dta},\cite{Dehghani:2019xjc},\cite{Dehghani:2021civ},\cite{Dehghani:2022xac},\cite{Dehghani:2021cyb}, arcsin-electrodynamics\cite{Kruglov_2015}, and Born-Infeld theory\cite{Born1,Born2}, each addressing certain shortcomings of Maxwell's original theory. The nonlinear electrodynamics framework provides more comprehensive insights into high-magnetization scenarios found in neutron stars and pulsars\cite{Bialynicka-Birula}\cite{H. J. Mosquera}. Furthermore, these theories effectively eliminate both the singularities associated with the Big Bang and those found in black holes by altering the geometry of spacetime.\\
Gravity in AdS space-time is intriguing due to the holographic principle. By utilizing this correspondence, we can study the behavior of its dual field theory. Some notable topics include quark-gluon plasma, phase transitions, condensed matter systems and transport coefficients.\\
Our aim is to investigate the field theory dual associated with nonlinear Yang-Mills gauge theory via the AdS/CFT correspondence\cite{Maldacena:1997re}. This necessitates the computation of transport coefficients, such as conductivity and the ratio of shear viscosity to entropy density ratio. Hydrodynamics serves as an effective theoretical framework for characterizing near-equilibrium phenomena occurring over large distances and timescales\cite{Landau,Son,Policastro2001}. The foundational conservation equations, together with constitutive relations, lead to the formulation of hydrodynamic equations through a gradient expansion around equilibrium, which can be expressed as follows:
\begin{align}
	& \nabla_{\mu} J^{\mu} = 0, \quad \nabla_{\mu} T^{\mu\nu} = 0,\\ & J^{\mu} = n , u^{\mu} - \sigma T P^{\mu\nu} \partial_{\nu}\left(\frac{\mu}{T}\right), \nonumber\\ & T^{\mu\nu} = (\rho + p) u^{\mu} u^{\nu} + p g^{\mu\nu} - \sigma^{\mu\nu}, \nonumber\\ & \sigma^{\mu\nu} = P^{\mu \alpha} P^{\nu \beta} \left[\eta(\nabla_{\alpha} u_{\beta} + \nabla_{\beta} u_{\alpha}) + \left(\zeta - \frac{2}{3}\eta\right) g_{\alpha\beta} \nabla \cdot u\right], \nonumber\\ & P^{\mu\nu} = g^{\mu\nu} + u^{\mu} u^{\nu}, \nonumber
\end{align}
 where $n$, $\eta$, $\zeta $, $\sigma ^{\mu \nu }$ , $\sigma$ and $P^{\mu \nu }$ are charge density, shear viscosity, bulk viscosity, shear tensor, conductivity and projection operator, respectively \cite{Kovtun2012}. $\eta$, $\zeta $ and $\sigma $ are known as transport coefficients denote charge density, shear viscosity, bulk viscosity, shear tensor, conductivity, and projection operator, respectively. The parameters. $\eta$, $\zeta $, $\sigma ^{\mu \nu }$ are categorized as transport coefficients.\\
The fluid-gravity correspondence, as a specific instance of the gauge-gravity duality, asserts that black holes within AdS spacetime correspond to stationary solutions of the relativistic hydrodynamics equations\cite{Son}-\cite{J.Bhattacharya2014}. The conductivity can be derived using both the Green-Kubo formula and the Witten prescription:
\begin{equation} \label{kubo}
	\sigma^{ij}_{(ab)}(k{\mu}) = -\lim_{\omega \to 0} \frac{1}{\omega} \Im G^{ij}_{(ab)}(k{\mu}),
\end{equation}
where the indices $a,b$ correspond to the $SU(2)$ group symmetry, while  $i,j$ indices refer to  $SU(2)$ refer to translational symmetry across the spatial dimensions $x,y$. \\
 In the context of the conformal field theory (CFT), there exists no distinction between upper and lower indices due to the flat Minkowski metric characterizing the boundary theory. A lower bound for direct current conductivity exists, defined as 
$\sigma \geq \frac{1}{e^2}=1$, where $e$ symbolizes the charge of the gauge field\cite{Grozdanov:2015qia,Grozdanov:2015djs}. However, it is important to note that  $e$ does not represent a unit of charge within the boundary theory. This bound has been realized in cases such as graphene\cite{Ziegler} but is often violated in situations involving massive gravity\cite{Baggioli:2016oqk}, models under abelian conditions\cite{Baggioli:2016oju}, in non-abelian Born-Infeld theory\cite{Sadeghi:2021qou} and non-abelian exponential guage theory of Yang-Mills type\cite{Sadeghi:2024ifg}. In this paper, we aim to explore the impact of non-abelian fractional gauge theory as a nonlinear electrodynamics framework on the conductivity bound of its strongly coupled dual theory.\\
%--------------------------------------------------------------------------
\section{  AdS black brane coupled to nonlinear gauge theory Solution}
\label{sec2}

\indent The action for the four-dimensional non-abelian fractional gauge theory with a negative cosmological constant is given as follows  \cite{Kruglov}\cite{KRUGLOV2015299},\cite{Kar:2024zbo},
\begin{eqnarray}\label{action}
	S=\int d^{4}  x\sqrt{-g} \bigg[\frac{1}{16\pi G}(R-2\Lambda) -\frac{\mathcal{F}}{4\pi(1+2 q_1 \mathcal{F})^2}\bigg],
\end{eqnarray}
where $R$ denotes the Ricci scalar, $l$ represents the AdS radius, and, $\mathcal{F}={\bf{Tr}}( F_{\mu \nu }^{(a)} F^{(a)\, \, \mu \nu })$ is the Yang-Mills invariant quantity, with $q_1$ being the nonlinear coupling constant. The trace is defined as ${\bf{Tr}}(.)=\sum_{i=1}^{3}(.).$ \\
$F_{\mu \nu }^{(a)}$ is the Yang-Mills  The field strength tensor for the $SU(2)$ Yang-Mills field is expressed as:
\begin{align} \label{YM}
F_{\mu \nu } =\partial _{\mu } A_{\nu } -\partial _{\nu } A_{\mu } -i[A_{\mu }, A_{\nu }],
\end{align}
The gauge field can be represented using the diagonal generator of the Cartan subalgebra of $SU(2)$.\\
AS $q_1 \to 0 $, the nonlinear term asymptotically becomes the standard linear non-abelian Yang-Mills term  \cite{Shepherd:2015dse}.\\
The equations of motion are derived by varying the action in \ref{action} with respect to  $g_{\mu \nu } $ and $A^{(a)}_{\nu}$,
\begin{eqnarray}\label{eom}
	&&   \frac{1}{16 G \pi}(R_{\alpha \beta }+\Lambda g_{\alpha \beta }-\frac{1}{2}g_{\alpha \beta } R)   + \frac{2 q_1 F_{\alpha }{}^{\gamma } F_{\beta \gamma } \mathcal{F}}{\pi \bigl(1 + 2 q_1 \mathcal{F}\bigr)^3}  -  \frac{F_{\alpha }{}^{\gamma } F_{\beta \gamma }}{2 \pi \bigl(1 + 2 q_1 \mathcal{F}\bigr)^2} \nonumber \\ 
	&& + \frac{\mathcal{F} g_{\alpha \beta}}{8 \pi \bigl(1 + 2 q_1 \mathcal{F}\bigr)^2}=0,
\end{eqnarray}
\begin{eqnarray}\label{EOM2}
\nabla^{\mu}\Bigg( \frac{F_{\mu \nu }}{\bigl(1 + 2 q_1 \mathcal{F} \bigr)^2}-\frac{4 q_1 \mathcal{F} F_{\mu \nu }}{
		\bigl(1 + 2 q_1 \mathcal{F}\bigr)^3}  \Bigg)=0
\end{eqnarray}
Our objective is to discover an asymptotically AdS black brane solution within a four-dimensional flat symmetric spacetime. We begin by considering the following ansatz:
\begin{equation}\label{metric1}
ds^{2} =-f(r)dt^{2} +\frac{dr^{2} }{f(r)} +\frac{r^2}{l^2}(dx^2+dy^2),
\end{equation}
where $f(r)$ is the metric function to be determined, and $l$ is a constant associated with the cosmological constant.\\
Next, we express the gauge field in terms of the matrix $H_1= \frac{i}{\sqrt{2}}\begin{pmatrix}1 & 0 \\ 0 & -1\end{pmatrix}$, which serves as the diagonal generator of the Cartan subalgebra of 
$SU(2)$ \cite{Shepherd:2015dse}. The gauge field is then written as:
\begin{equation}\label{background}
{\bf{A}}^{(a)} =\frac{i}{\sqrt{2}}h(r)dt\begin{pmatrix}1 & 0 \\ 0 & -1\end{pmatrix}.
\end{equation}
Considering Equations \ref{EOM2} and \ref{metric1}, we can reformulate Eq. \ref{EOM2} into the following expression:
\begin{equation}
\frac{h''}{2}+\frac{h'}{r}-\frac{8 q_1 h'^3}{r \left(4 q_1 h'^2+1\right)}+\frac{96 q_1^2 h'^4 h''}{ \left(4 q_1 h'^2+1\right)^2}-\frac{20 q_1 h'^2
	h''}{\left(4 q_1 h'^2+1\right)}=0,
\end{equation}
This differential equation can be solved to obtain the function $h(r)$,\\
\begin{equation}\label{h}
h(r)=\mu + \int^{r} D(u)du,
\end{equation}
where,
\begin{equation}
	D(u)=\text{Root}\left[64 q_1^3 x^6+48 q_1^2 x^4+4 e^{c_1} q_1 u^2 x^3+12
	q_1 x^2-e^{c_1} u^2 x+1,x\right].
\end{equation}
Here, $\mu$ is an integration constant referred to as the chemical potential of the quantum field at the AdS boundary. The value of $\mu$ can be determined using the regularity condition at the horizon, specifically $A_t(r_h) = 0$ \cite{Dey:2015poa}. This gives us:
\begin{equation}
\mu=-\int^{r_h} D(u)du,
\end{equation}
\noindent the invariant scalar $\mathcal{F}_{YM}={\bf{Tr}}( F_{\mu \nu }^{(a)} F^{(a)\, \, \mu \nu })$ for the fields is defined as:
\begin{equation}
  F_{tr}^{(a)} =-F_{rt}^{(a)}=\frac{-i}{ \sqrt{2}}D(r)\begin{pmatrix}1 & 0 \\ 0 & -1\end{pmatrix},
\end{equation}
By setting $q_1=0$, our findings reduce to the non-abelian Yang-Mills solution.\\
The satisfaction of the Bianchi identity is expressed as follows:\\
\begin{equation}
\nabla_{\alpha }F^{(a)}_{\mu \nu}+\nabla_{\nu }F^{(a)}_{\alpha \mu}+\nabla_{\mu }F^{(a)}_{ \nu \alpha}=0.
\end{equation}
Now, when we examine the $tt$ component of Eq. (\ref{eom}), we have:
\begin{align}\label{tt}
	&2 r f(r)^2 h'(r)^2 \left(4 q_1 r h'(r)^2+r\right)-4 \pi  f(r)^3 \left(4 q_1 h'(r)^2-1\right)^3 \nonumber\\&-4 \pi  r f(r)^2 f'(r) \left(1-4 q_1 h'(r)^2\right)^2 \left(4 q_1 h'(r)^2-1\right)\nonumber\\&-r^2f(r)^2 \left(4 q_1 h'(r)^2-1\right) \left(64 \pi  \Lambda  q_1^2 h'(r)^4-(32 \pi  \Lambda  q_1+1)
	h'(r)^2+4 \pi  \Lambda \right)=0,
\end{align} 
from Eq. (\ref{eom}), we can derive $f(r)$ by solving:
\begin{align}\label{sol}
&f(r)=-\frac{2m}{r}\nonumber\\&-\frac{1}{r}\int^r\frac{u^2 \left(-16 q_1 (G-3 \Lambda  q_1) h'(u)^4+12 (G+\Lambda  q_1) h'(u)^2+64 \Lambda  q_1^3
	h'(u)^6+\Lambda \right)}{\left(4 q_1 h'(u)^2+1\right)^3}du,
\end{align}
where $m$ is an integration constant. Note that $r$ is the radial coordinate that transitions us from the bulk to the boundary.\\
The event horizon is situated at the point where $f(r_h)=0$, and the value of $m$ is determined by this condition.
\begin{align}
	&f(r)=2m(\frac{1}{r_h}-\frac{1}{r})\nonumber\\&-\frac{1}{r}\int^r_{r_h}\frac{u^2 \left(-16 q_1 (G-3 \Lambda  q_1) h'(u)^4+12 (G+\Lambda  q_1) h'(u)^2+64 \Lambda  q_1^3
		h'(u)^6+\Lambda \right)}{\left(4 q_1 h'(u)^2+1\right)^3}du,
\end{align}
The Hawking temperature for this black brane is given by:
  \begin{align}\label{Temp}
  T& =\frac{f'(r_h)}{4 \pi}=\frac{2m}{4 \pi r_h^2}\nonumber\\&-\frac{r_h \left(-16 q_1 (G-3 \Lambda  q_1) h'(r_h)^4+12 (G+\Lambda  q_1) h'(r_h)^2+64 \Lambda  q_1^3
  	h'(r_h)^6+\Lambda \right)}{4 \pi \left(4 q_1 h'(r_h)^2+1\right)^3}
  \end{align}
\section{Color non-abelian DC conductivity}
\label{sec3}
\indent To calculate the retarded Green's function $ <J^{\mu}(\omega)J^{\nu}(-\omega)>$, we perturb the gauge field as $A_{\mu} \to A_{\mu}+\tilde{A}_{\mu}$ and expand the action up to second order in the perturbed part $\tilde{A}_{\mu}$ \cite{Baggioli:2016oju}-\cite{Policastro2002}. We can identify the current $J_a^{\mu}$ from the perturbed part of the action as follows:
 \begin{equation}
 S \to S +\int d^4x\,\, A^a_{\mu}J_a^{\mu}.
 \end{equation}
The retarded Green's function can be obtained from the AdS/CFT correspondence as:
\begin{equation}
G^{(ab)}_{\mu \nu}(k_{\mu})=<J^{(a)}_{\mu}(\omega)J^{(b)}_{\nu}(-\omega)>=\frac{\delta^2 S}{\delta \tilde{A}^{0\,\,\mu}_{(a)} \delta \tilde{A}^{0\,\,\nu}_{(b)}},
\end{equation}
where $\tilde{A}^0_{\nu}$ denotes the value of the gauge field at the boundary. We can compute the non-abelian color DC conductivity using the Green-Kubo formula:
\begin{equation} \label{kubo2}
\sigma^{(ab)}_{\mu \nu}(k_{\mu})=-\mathop{\lim }\limits_{\omega \to 0} \frac{1}{\omega } \Im G^{(ab)}_{\mu \nu}(k_{\mu}),
\end{equation}
indicating that the DC conductivities are related to the retarded Green's function of the boundary current $\mathop{J}^i$. The conductivities along the 
$i$ and $j$  directions can be expressed as:
\begin{equation}
\sigma^{(ab)}_{\mu \nu}(k_{\mu})(x,y) = \frac{\delta J^{\mu}_{(a)}(x)}{\delta E^{\nu}_{(b)}(y)}.
\end{equation}
Since the gauge field theory has only the zeroth component, and this component vanishes on the boundary, the rotational symmetry is preserved on the boundary. Due to the rotational symmetry in the $xy$ plane, the color DC conductivity is a scalar quantity.
\begin{equation}
	\sigma^{ij}_{(ab)}=\sigma_{(ab)}\delta^{ij}.
\end{equation}
Next, we consider the perturbative part of the gauge field as  $\tilde{A}_x=\tilde{A}_x(r)e^{-i\omega t}$ where $\omega$ is small, ensuring that we are in the hydrodynamic regime. By inserting the perturbed gauge field into the action (Eq. (\ref{action})) and retaining terms up to second order in $\tilde{A}$, we obtain:
\begin{align}\label{action-2}
S^{(2)}&=\int d^4x \frac{1}{\pi  r^2 f
	\left(4 q_1 h'^2+1\right)^3} \left[\left(4 q_1 h'^2-1\right) \left((\partial_r\tilde{A}_x^{(1)})^2+(\partial_r\tilde{A}_x^{(2)})^2+(\partial_r\tilde{A}_x^{(3)})^2\right)\right. \nonumber\\
&\left.-\Big((\tilde{A}_x^{(1)})^2+(\tilde{A}_x^{(2)})^2+\tilde{A}_x^{(3)})^2\Big)\omega ^2 + 2\Big(\tilde{A}_x^{(1)})^2+\tilde{A}_x^{(2)}\Big)\right].
\end{align}
By variation of the action $S^{(2)}$ with respect to $\tilde{A}_x ^a$ we have,
\begin{align}\label{PerA1}
	&64 q_1 r f^2 \tilde{A}_x^{(1)'} h' h'' \left(2 q_1 h'^2-1\right) \nonumber\\&-2\left(16 q_1^2 h'^4-1\right) \left(f
	\left(\tilde{A}_x^{(1)'} \left(r f'-2 f\right)+r f \tilde{A}_x^{(1)''}\right)+r \tilde{A}_x^{(1)} \left(2 h^2+\omega ^2\right)\right)=0,
\end{align}
\begin{align}\label{PerA2}
	&64 q_1 r f^2 \tilde{A}_x^{(2)'} h' h'' \left(2 q_1 h'^2-1\right) \nonumber\\&-2\left(16 q_1^2 h'^4-1\right) \left(f
	\left(\tilde{A}_x^{(2)'} \left(r f'-2 f\right)+r f \tilde{A}_x^{(2)''}\right)+r \tilde{A}_x^{(2)} \left(2 h^2+\omega ^2\right)\right)=0,
\end{align}
\begin{align}\label{PerA3}
	&64 q_1 r f^2 \tilde{A}_x^{(3)'} h' h'' \left(2 q_1 h'^2-1\right)\nonumber\\&-2\left(16 q_1^2 h'^4-1\right) \left(f
	\left(\tilde{A}_x^{(3)'} \left(r f'-2 f\right)+r f \tilde{A}_x^{(3)''}\right)+2 h^2 r \tilde{A}_x^{(3)} \right)=0,
\end{align}
To begin with, we address Eq. (\ref{PerA1}), Eq. (\ref{PerA2}), and Eq. (\ref{PerA3}) in proximity to the event horizon. By expanding these equations near the horizon and utilizing the relation $f \sim 4\pi T(r-r_h)$, derived from the definition of Hawking temperature, we can express the solution for $\tilde{A}_x^{(a)}$ as follows,
\begin{align}
\tilde{A}_x^{(a)}\sim (r-r_h)^{z_a} \, , \qquad a=1,2,3\,\,\,\,,   
\end{align}
where,
\begin{align}\label{z12}
z_1&=z_2=\pm i \frac{\sqrt{2 h(r_h)^2+\omega ^2}}{4 \pi T} , \\
\label{z3}
z_3&=\pm i \frac{\omega }{4 \pi T}.
\end{align}
To solve Eq. (\ref{PerA1}), Eq. (\ref{PerA2}), and Eq. (\ref{PerA3}), we employ the following ansätze:
\begin{align}\label{EOMA1}
\tilde{A}_x^{(1)}=\tilde{A}^{(1)}_{\infty}\Big(\frac{-3}{\Lambda r^2}f\Big)^{z_1}\Big(1+i\omega h_1(r)+\cdots\Big) ,
\end{align}
\begin{align}\label{EOMA2}
\tilde{A}_x^{(2)}=\tilde{A}^{(2)}_{\infty}\Big(\frac{-3}{\Lambda r^2}f\Big)^{z_2}\Big(1+i\omega h_2(r)+\cdots\Big) ,
\end{align}
\begin{align}\label{EOMA3}
\tilde{A}_x^{(3)}=\tilde{A}^{(3)}_{\infty}\Big(\frac{-3}{\Lambda r^2}f\Big)^{z_3}\Big(1+i\omega h_3(r)+\cdots\Big) ,
\end{align}
where $\tilde{A}^{(a)}_{\infty}$ denotes the field values at the boundary, and $z_i$ corresponds to the negative signs in equations \ref{z12} and \ref{z3}.\\
Next, by substituting Eq. \ref{EOMA3} into Eq. (\ref{PerA3}) and focusing on the first order of $\omega$ we arrive at the following expression:
\begin{align}\label{h3}
	&-2 r \left(16 q_1^2 h'^4-1\right) \left(f'(r) \left(r h_3'-4\right)+r f''\right)+64 q_1 r^2 f' h' h''
	\left(2 q_1 h'^2-1\right)\nonumber\\&+2 f \left(r^2 h_3'' \left(1-16 q_1^2 h'(r)^4\right)+2 \left(r h_3'-3\right)
	\left(16 q_1^2 h'^4-1\right)+32 q_1 r h' h'' \left(r h_3'-2\right) \left(2 q_1 h'^2-1\right)\right)\nonumber\\&=0,
\end{align}
The form of the solution for $h_3(r)$  is given by:
\begin{equation}
 h_3(r) =C_3 + C_4\int^r \frac{u^2 (1+4 q_1 h'(u)^2 )^3}{f(u)(1-4 q_1 h'(u)^2 )}du- \int^r \frac{u f'(u)-2 f(u)}{f(u) u
	}du,
\end{equation}
where $C_3$ and $C_4$ are constants of integration. Near the horizon, the behavior of $h_3(r)$ can be approximated as:
\begin{equation}\label{C4}
h_3 \approx  \bigg(\frac{C_4 r_h^2 (1+4 q_1 h'(r_h)^2 )^3}{4 \pi T(1-4 q_1 h'(r_h)^2 )}-1\bigg) \log(r-r_h)+\text{finite terms}.
\end{equation}
The value of  $C_4$ is determined by ensuring regularity at the horizon, resulting in:
\begin{equation}
C_4=\frac{4 \pi T(1-4 q_1 h'(r_h)^2 ) }{ r_h^2 (1+4 q_1 h'(r_h)^2 )^3} .
\end{equation}
By substituting the solution for $\tilde{A}^{(3)}_{x}$ into Eq. (\ref{action-2}) and varying with respect to $\tilde{A}^{(3)}_{\infty}$, we can extract the Green's function:
\begin{equation} \label{Green1}
 G_{xx}^{(33)} (\omega ,\vec{0})=\tilde{A}_x^{(3)})^{*}f(r) \partial _{r}\tilde{A}_{x}^{(3)}\bigg|_{r \to \infty}=-i \frac{\omega r_h^2 C_4}{4\pi T},
\end{equation}
substituting $f(r)$, $h_3(r)$ and $h'(r)$ into this equation, along with utilizing Eq. (\ref{kubo2}), we find:  
\begin{eqnarray}\label{sigma33}
\sigma_{xx}^{(33)}=-\mathop{\lim }\limits_{\omega \to 0} \frac{1}{\omega } \Im G^{ij}(k_{\mu}) =\frac{1-4 q_1 h'(r_h)^2}{(1+4 q_1 h'(r_h)^2)^3},
\end{eqnarray}
The conductivity bound is violated for this model.\\
In the limit of $q_1 \to 0$  we have non-abelian Yang-Mills theory and the conductivity bound is saturated\cite{S Parvizi}.
\begin{eqnarray}
\sigma_{xx}^{(33)} =1.
\end{eqnarray}
The values of $\sigma_{xx}^{(11)}$ and  $\sigma_{xx}^{(22)}$ can be determined using the same method applied to $\sigma_{xx}^{(33)}$, resulting in:
\begin{eqnarray}
\sigma_{xx}^{(11)} =\sigma_{xx}^{(22)} =0.
\end{eqnarray}
This suggests that the color non-abelian DC conductivity, in relation to color indices, takes on a diagonal form, and Ohm's law similarly retains a diagonal structure within this context. Furthermore, if we examine the gauge field as outlined in \ref{background} along the 
$\sigma_{1}$ or $\sigma_{2}$ directions, the conductivity in those directions will be non-zero.\\
Since this is the Einstein-Hilbert theory in the presence of non-linear Yang-Mills terms, the shear viscosity to entropy density ratio in this model is equal to $\frac{1}{4\pi}$, or in other words, the KSS \footnote{Kovtun-Son-Starinet} conjecture is saturated\cite{Sadeghi:2024ifg}. The KSS bound is violated in the massive gravity \cite{Sadeghi:2019muh,Sadeghi:2020lfe} and Gauss-Bonnet gravity\cite{Sadeghi:2022kgi,Sadeghi:2015vaa} and arcsin ads black brane\cite{Sadeghi:2024ajb}.\\
%--------------------------------------------------------------------------
 \section{Conclusion}
\noindent We have introduced the Einstein non-abelian nonlinear gauge theory and explored its AdS black brane solution, calculating the non-abelian color DC conductivity associated with this model. It has been conjectured that the conductivity is universally bounded by the value $\sigma \geq 1$ {\footnote{We set $\frac{1}{e^2}=1$.}}. Our findings indicate that this conductivity bound is violated in our model, although it is saturated in the case of Yang-Mills theory \cite{S Parvizi}.\\
The violation of the conductivity bound may occur in nonlinear models, influenced by the specific formulation of the nonlinear theory or the coupling of $F^2$ to the scalar field. We interpret this violation as being comparable to quark-quark coupling, suggesting a relationship with Mott insulators.\\
We conclude that the conductivity corresponds to the derivative of the Lagrangian with respect to $F^2$, evaluated at the event horizon, and is saturated for Yang-Mills theory, though it may be violated in a nonlinear context.\\
Since the bound on the conductivity is greater than or equal to the inverse square of the charge carried by the gauge field, we conclude that the charge carried by the gauge field in this model is different from the charge carried by the Yang-Mills theory.\\
%Additionally, we find that the ratio of shear viscosity to entropy density for our model is given by $\frac{\eta }{s}=\frac{1 }{4 \pi}$. This indicates that the Kovtun-Son-Starinets (KSS) bound \cite{Kovtun:2004de}, which states this value for Einstein-Hilbert gravity, is upheld in our model as well. Notably, the ratio 
%$\frac{\eta }{s}$ is proportional to the inverse square of the coupling in the field theory, expressed as $\frac{\eta }{s} \sim \frac{1 }{\lambda^2}$. This implies that the coupling in the field theory dual to our model matches that of the theory dual to the Einstein AdS black brane solution, while the color conductivity differs.
%--------------------------------------------------------------------------
%\vspace{1cm}
%\noindent {\large {\bf Acknowledgment} } We would like to thank ............. for useful comments and suggestions. We also thank the referees of CJP for the valuable comments which helped us to improve the manuscript.
%--------------------------------------------------------------------------

\vspace{1cm}
\noindent {\large {\bf Data Availability } } Data generated or analyzed during this study are provided in full within the published article.
%--------------------------------------------------------------------------

\vspace{1cm}
\noindent {\large {\bf Competing interests }  The author declares there are no competing interests.
%--------------------------------------------------------------------------

\end{document}